\documentstyle{l-aa}

\newcommand{\apj}{{ApJ\ }}
\newcommand{\aap}{{A\&A\ }}

\newcommand{\aj}{{AJ\ }}
\newcommand{\apjs}{{ApJS\ }}
\newcommand{\apjl}{{ApJ Lett.\ }}

\begin{document}

\thesaurus{03(11.02.2; 13.07.2)}

\title{TeV gamma-ray observations of three X-ray selected BL Lacs}

\author{ M.D. Roberts\inst{1,2}, P. McGee\inst{2},
  S.A. Dazeley\inst{2}, P.G. Edwards\inst{3},
  T. Hara\inst{4},
  J. Holder\inst{1},
  A. Kawachi\inst{1}, T. Kifune\inst{1},
  Y. Matsubara\inst{7},
  Y. Mizumoto\inst{8},
  M. Mori\inst{1}, H. Muraishi\inst{6},
  Y. Muraki\inst{7}, T. Naito\inst{8}, 
  K. Nishijima\inst{10},
  S. Ogio\inst{5}, T. Osaki\inst{5},
  J.R. Patterson\inst{2},
  G.P. Rowell\inst{1,2}, T. Sako\inst{7},
  K. Sakurazawa\inst{5}, R. Susukita\inst{11},
  T. Tamura\inst{12}, T. Tanimori\inst{5},
  G.J. Thornton\inst{2}, S. Yanagita\inst{6},
  T. Yoshida\inst{6} and  T. Yoshikoshi\inst{1}}

\date{Received 7 September 1998; Accepted 1 December 1998}

\offprints{roberts@icrr.u-tokyo.ac.jp}

\institute{Institute for Cosmic Ray Research, University of Tokyo,
  Tokyo 188, Japan
 \and Department of Physics and Mathematical Physics, University of
  Adelaide, South Australia 5005, Australia
 \and Institute of Space and Astronautical Science, Kanagawa 229,
  Japan
 \and Faculty of Commercial Science, Yamanashi Gakuin University,
  Yamanashi 400, Japan
 \and Department of Physics, Tokyo Institute of Technology,
  Tokyo 152, Japan
 \and Faculty of Science, Ibaraki University, Ibaraki 310, Japan
 \and Solar-Terrestrial Environment Laboratory, Nagoya University,
  Aichi 464, Japan
 \and National Astronomical Observatory of Japan, Tokyo 181, Japan
 \and Faculty of Education, Miyagi University of Education,
  Miyagi 980, Japan
 \and Department of Physics, Tokai University, Kanagawa 259,
  Japan
 \and Institute of Physical and Chemical Research, Saitama 351-01,
  Japan
 \and Faculty of Engineering, Kanagawa University, Kanagawa 221,
  Japan}

\maketitle
\markboth{Roberts et al.}{TeV Observations of 3 X-ray selected BL~Lacs}
\begin{abstract}

Despite extensive surveys of extragalactic TeV gamma-ray candidates
only 3 sources have so far been detected. All three are northern
hemisphere objects and all three are low-redshift X-ray selected
BL~Lacs (XBLs). In this paper we present the results of observations
of the three nearest southern hemisphere XBLs (PKS0548$-$322, PKS2005$-$489
and PKS2155$-$304) with the CANGAROO 3.8m imaging telescope.
 During the period of observation we estimate that
the threshold of the 3.8\,m telescope was $\sim$1.5\,TeV. Searches for
both steady and short timescale emission have been performed for
each source. Additionally, we are able to monitor the X-ray state of each
source on a daily basis and we have made contemporaneous measurements
of optical activity for PKS0548$-$322 and PKS2155$-$304.


\end{abstract}

\section{Introduction}

At the highest photon energies observable by satellite borne detectors, the
bulk of detectable sources are active galaxies of the blazar class. These,
and similar objects, make a natural sample for observation by ground based
atmospheric Cherenkov detectors. 
At present only three extragalactic sources have been found, from Cherenkov
imaging observations, to emit gamma-rays  
above 300\,GeV: Mkn~421 (\cite{pun92}, \cite{mac95}), 
Mkn~501 (\cite{qui97}, \cite{aha97}) and 
1ES~2344+514 (\cite{cat97a}).
All three sources are nearby (z$<$0.05) BL~Lacs and, based on their 
log~($F_X$/$F_r$) ratios, are classified as XBLs.

The $\nu$F$_{\nu}$ spectral energy distributions (SEDs) of blazars show a 
characteristic ``double humped''
structure which is naturally explained in terms of a
synchrotron self-Compton (SSC) mechanism, with
the lower energy peak due to synchrotron emission, and the higher energy
peak due to Compton scattering of the synchrotron photons 
from the same electron population
(see e.g. \cite{mar92}). 
Other mechanisms, including different sources of seed photons for the
Compton process (\cite{der94,sik94}) and the
 possibility of production of gamma-rays
from protons in the jet (\cite{man93}), have also been proposed. 
Comparisons of SEDs between XBLs and radio selected BL~Lacs (RBLs)
suggest that XBLs may have more compact emission regions, with higher
electron energies and stronger magnetic fields than RBLs (\cite{sam96}).
An important feature of BL~Lacs is their strong variability at all
wavelengths. At TeV energies Mkn~421 has been seen to increase in
luminosity by a factor of more than 10 in less than one hour (\cite{mce97}).

Multiwavelength studies have shown that the X-ray luminosity of XBLs is
most strongly correlated with the changes in TeV luminosity.
Based on X-ray measurements \cite{ste96} have produced a catalogue 
of XBLs that are most likely to produce detectable levels of TeV emission. 
Assuming an SSC mechanism \cite{ste96} note that the Compton
and synchrotron emission spectra have similar shapes, but that the Compton
component is upshifted in energy by the square of the Lorentz factor of
the electrons in the jet. Under the further assumption that the
Compton and synchrotron luminosities are nearly equal for XBLs 
(as is the case for Mkn~421 and PKS2155$-$304), \cite{ste96} derive
the following relationship between X-ray and TeV fluxes : 

\begin{equation}
\nu_{TeV}F_{TeV} \sim \nu_{x}F_{x}
\label{Stecker}
\end{equation}

{\noindent}where $\nu$ is the frequency and $F$ the flux in each energy band.
Another important consideration for estimating detectability at
TeV energies is the proximity of the source as the gamma-rays
interact via pair production with intergalactic soft photon fields. At TeV
energies most absorption occurs due to interaction with infra-red photons.
\cite{ste96} have estimated the attenuation
of the TeV signal based on models of the IR background published in
 \cite{ste97}. 
Recent observations of Mkn~421 and Mkn~501 (\cite{mce97,aha97,qui97}), 
however, show that
the TeV emission spectrum is unmodified up to energies of at least 5 TeV,
indicating that the density of the IR background assumed 
in \cite{ste97} may be
too high. A more recent estimation of the opacity of the IR background
has been made in \cite{ste98}. 

Prior to 1997 the CANGAROO 3.8m telescope was used to observe a number
of extragalactic objects, including RBLs and XBLs (\cite{rob98} 1998). 
Following
the success of TeV observations of Mkn~421 and Mkn~501 in conjunction
with multiwavelength campaigns, we have concentrated observations
on nearby XBLs. We have paid special attention to contemporaneous 
monitoring of these highly
variable sources, where possible, at optical and X-ray wavelengths.
Presented here are observations of the three 
most promising southern hemisphere 
TeV candidates from \cite{ste96} : PKS0548$-$322, 
PKS2005$-$489 and PKS2155$-$304.


\section{The CANGAROO 3.8m telescope}

The CANGAROO 3.8m imaging telescope is located near Woomera, South Australia
(longitude $137^{\circ}47'$E, latitude $31^{\circ}06'$S, 160m a.s.l).
The telescope consists of a single 3.8m diameter parabolic reflector
with a 3.8m focal length. A high resolution imaging camera is located at the
prime focus consisting of 256 Hamamatsu R2248 photomultiplier tubes
arranged in a $16 \times 16$ grid. 
 The photomultipliers are separated by $0.18^{\circ}$, giving
a total field of view (side-side) of $2.9^{\circ}$. The photo-cathode of
each tube subtends $0.12^{\circ} \times 0.12^{\circ}$ representing  40\%
of the total field of view.

An event trigger is generated when 3-5 phototubes exceed the
discriminator threshold (which is estimated to be 3 photo-electrons).
Under this triggering condition the current gamma-ray
energy threshold of the 3.8m telescope is estimated to be 1.5 TeV, and the
vertical trigger rate due to background cosmic rays is around 2Hz.
More detailed descriptions of the CANGAROO 3.8m telescope can be found in
\cite{har93,yos97} and \cite{rob98} (1998).

As well as TeV observations we also have roughly contemporaneous X-ray and
optical (for PKS2155$-$304 and PKS0548$-$322) data available. Optical 
observations were made
from the Woomera site, using a Celestron C14 Schmidt-Cassegrain telescope
with an Optec UBVRI filter set and SBIG ST-6 CCD.
Exposures were 180 seconds in duration with the start times of time-series  
recorded from a  UT clock ($\pm$1 second); timing 
within time-series runs was from
a PC clock (maximum duration about 2 hours) with, at most, only a 
few seconds of expected error. The errors in optical flux are around
1\% --- much smaller than the flux variations that have been observed
from these sources.

We are able to monitor the X-ray state of each source via daily
observations by the Rossi X-ray timing explorer satellite (RXTE). The
all sky monitor (ASM) on this satellite provides nearly continuous X-ray
coverage of the whole sky in the energy range 2-10keV. 

\section{Data analysis}
\label{data_anal}

Prior to image analysis the arrival times of the raw events are binned into
one minute intervals and the rate distribution is checked for the
presence of cloud or any electronics problems. ADCs and TDCs are calibrated
using LED ``flasher'' data taken at the beginning of each observation.
The phototubes that are included in the image are selected on the
following criteria:

1: The TDC of the tube must have fired (tube must have exceeded
discrimination threshold)

2: The ADC signal in the tube must be at least 1 SD above the RMS
of background noise for that tube

{\noindent}An image is considered suitable for parameterization if it contains at least
5 tube signals, and if the total signal for all tubes in the image 
exceeds 150 ADC counts (around 20 p.e.). The images are parameterized as
a simple ellipse, after the method of \cite{hil85}. The parameters
used and their gamma-ray selection domains are :
\\
\\
$0.5^{\circ} <$ Distance $< 1.2^{\circ}$\\
$0.01^{\circ} <$ Width $< 0.1^{\circ}$\\
$0.1^{\circ}<$  Length $< 0.35^{\circ}$\\
alpha $< 10^{\circ}$\\
\\
The gamma-ray parameter domains have been optimized on 
Monte Carlo simulations of the response of the CANGAROO 3.8m telescope
to both gamma-ray and proton initiated EAS. 
We estimate that our image selection rejects $\sim$99\% of the cosmic ray
background while retaining $\sim$40\% of the gamma-ray signal.
For more details of the image selection used for
the CANGAROO 3.8m see \cite{yos97} and \cite{rob98} (1998).

The set of total observations for each source have been tested for the
presence of a gamma-ray signals. The significances of the on-source
excesses have been estimated using a method based on that
suggested by \cite{li83}:

\begin{displaymath}
S  = \sqrt{2} \left\{ N_{on}  
{\rm ln} \left[\frac{1+\beta}{\beta} 
\left( \frac{N_{on}}{N_{on}+N_{off}} \right) \right] \right. 
\end{displaymath}

\begin{equation}\noindent 
 \ \ \ \ \ \ \ \ \ \ \  \left. + N_{off} {\rm ln} 
\left[ (1+\beta) 
\left( \frac{N_{off}}{N_{on}+N_{off}} \right) \right]  \right\}^{1/2}
\end{equation}

\noindent where S is the statistical significance and $\beta$ is the 
ratio of events in the on-source to off-source exposure. In practice
we estimate $\beta$ from the ratio of on-source to off-source events
in the range $30^{\circ} < {\rm alpha} < 90^{\circ}$ (where alpha
is the image orientation parameter) for all images that are 
considered suitable for parameterization. While the calculation
of S is not strictly correct, the error introduced by the uncertainty
in $\beta$ is small compared to the statistical uncertainties in $N_{on}$
and $N_{off}$.
As well as testing for DC emission, we have also tested for shorter
timescale gamma-ray emission. Each night's observation consists generally
of a matched pair of on and off-source runs
but where no matching off-source run
was collected, an off-source observation from a nearby night is used.
For both steady DC and night by night excess searches we calculate
upper limits on gamma-ray emission after \cite{pro84}.

For optical observations data reduction was performed via differential
 aperture photometry relative to the local field standards given in 
\cite{smi91}. Dark subtraction
 was performed during observation, and twilight flat-fields were applied
 during the reduction procedure.
The magnitude differences between the comparison stars in the 
PKS2155$-$304 and PKS0548$-$322 fields, obtained in the Woomera 
photometry, were compared with the differences found from 
\cite{smi91}. In all cases, the values agreed within error 
(VRI filters for both fields), but with the delta(I) value for the 
PKS0548$-$322 comparisons being at the limit of the errors. 
Thus, it was assumed that the differential magnitudes for the AGN 
could be converted directly to standard magnitudes by using the 
\cite{smi91} magnitudes for the comparison stars.
By observing the optical state of the AGN we are able to monitor the general
activity of the sources, and compare this activity to historical levels. 
While the correlation between optical and TeV luminosity seen in Mkn~421 
during short intense TeV flares has been relatively weak (\cite{mac95}),
 multiwavelength observations of PKS2155$-$304, for example, show flux
variations that are remarkably similar for wavelengths between infrared
and X-ray (\cite{ede95} 1995). 

The strong correlation between X-ray and TeV emission has been well 
documented (\cite{mac95,cat97b} 1997b), and we are able to monitor the X-ray
state of all three candidates via the ASM on the
RXTE. Adopting the method of \cite{cat97b} (1997b),
we convert the ASM counts (2-10 keV) from the quicklook
 analysis of each source to an X-ray flux by assuming that each candidate
source has a similar photon spectrum to the Crab Nebula (in the range 2-10
keV) and normalizing
the ASM data to the known flux of the Crab. 
Using the method suggested by \cite{ste96} (see Eq.~\ref{Stecker}) 
we can use the RXTE flux to predict the level of TeV emission from each source.
In this paper we will normalize the predicted TeV flux
using the TeV to RXTE(2-10keV) ratio measured by the Whipple collaboration
from Mkn~501 during the period $9^{th}$ - $15^{th}$ of April 1997
 (\cite{cat97b} 1997b). 
Extrapolation of fluxes from the Whipple
telescope energy threshold (350 GeV) to that of the CANGAROO telescope are
made assuming an integral photon spectrum of index $-$1.5. It is further
assumed that the photon energies of the candidates extend to at least 10 TeV.
We will assume an absorption of TeV gamma-rays in the cosmic IR based 
on the largest interaction length model from \cite{ste98}. 
This model 
predicts that the spectrum of a source such as Mkn~501 will not be
strongly affected below 30TeV, which is consistent with 
the most recent observations
from the HEGRA collaboration (\cite{kra98}). 

The ASM quicklook data provides an X-ray daily average from 
a number of ``dwells'' taken throughout each day, while 
the on-source TeV observations are made during
a typically 4-8 hour period each night. The X-ray and TeV fluxes of
XBLs are known to vary on timescales of less than hours, so care must be
taken when interpreting X-ray/TeV correlations from short-term flare activity.
Given this, and the relative insensitivity of the daily X-ray and TeV 
measurements, we will not present our data on timescales of less than
the typically 1-2 week TeV observation period of each source.

\section{PKS0548$-$322}

At a distance of z=0.069, PKS0548$-$322 is the closest known 
southern hemisphere
XBL (although it is still twice as distant as the nearest known northern XBL).
X-ray observations have shown that this object undergoes rapid changes
in both luminosity and spectral hardness. PKS0548$-$322 has not
been detected by the CGRO EGRET, but the X-ray brightness, flat X-ray
spectrum and proximity of this object make it a candidate for detection
at TeV energies.
CANGAROO observations of PKS0548$-$322 cover two new-moon periods
from the $28^{th}$ of October to the $27^{th}$ of November 1997. For this
period we have 9 nights of analyzable observations, yielding a total of
$\sim22$ hours of off-source and $\sim 26$ hours of on-source data. 
Analysis of this dataset shows no evidence for an on-source excess of
events in the gamma-ray selection domain. The significance of the on-source 
excess is $-1.1 \sigma$ with a corresponding $2 \sigma$ flux limit
above 1.5 TeV of $4.3 \times 10^{-12}$ photons cm$^{-2}$ s$^{-1}$. A night by
night search of the dataset shows no evidence for detectable flare
activity on the scale of $\sim$ 1 day.

X-ray measurements over the total CANGAROO observation period 
of PKS0548$-$322 show an average ASM count rate of
$0.14 \pm 0.07$, where the error is the quadrature error derived from the
error in each daily average X-ray count rate. For comparison the ASM count
rate for Mkn~501 for the same period is $0.946$, and $\sim~75$ for
the Crab Nebula. Based on the measured X-ray flux, and using the 
simple model described in section~\ref{data_anal} we predict that the
flux of gamma-rays above 1.5TeV should be $0.57 \times 10^{-12}$ 
photons cm$^{-2}$s$^{-1}$. An examination of the entire ASM data base for this
source from JD2450493 to JD2450863 shows an average ASM count rate of 
$0.14 \pm 0.02$. A
sliding window of the same length as the total duration of the CANGAROO
observations ($\sim 31$ days) shows that the maximum average X-ray flux within
this timescale in the ASM dataset is $0.45 \pm 0.08$. 

The optical activity of PKS0548$-$322 was monitored from Woomera on
9 consecutive nights from the 29$^{th}$ of October until the 6$^{th}$ of
November (excluding the $31^{st}$ of October). There is no evidence for any
variation between the 180s exposures for any night, and furthermore, no
evidence for any variability between the nights. The average magnitude 
measured in the R band (649nm) was $14.59 \pm 0.04$ (stat.) corresponding to
an average flux at this wavelength of 4.48 mJy, similar to other reported
optical measurements (see e.g \cite{xie96}).

The optical and X-ray measurements show that PKS0548$-$322 was at
a fairly typical level of activity during the period of the CANGAROO
observations.

\begin{figure}
 \picplace{70mm}
 \includegraphics{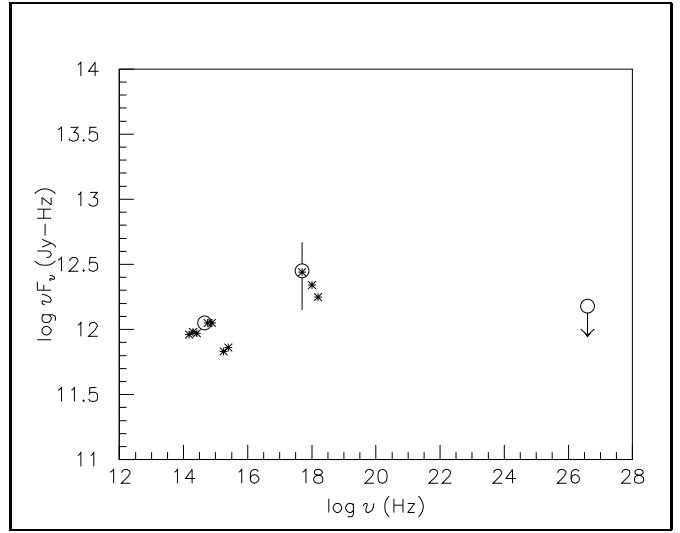}
   \caption{SED for PKS0548$-$322. The open circles indicate our measurements
and the contemporaneous RXTE flux
and the stars are non-contemporaneous measurements adapted from \cite{gho95}.}
 \label{pks0548}
\end{figure}

\section{PKS2005$-$489}
The radio to X-ray SED of PKS2005$-$489 (z=0.071) is very similar 
to that of the confirmed
TeV emitter Mkn~421 (\cite{sam95}). As with Mkn~421, X-ray observations
show that the X-ray spectrum of PKS2005$-$489 hardens with increasing
source intensity (\cite{gio90}).  
Although initially reported as a marginal EGRET detection (\cite{mon95}),
a more accurate background estimation decreased the
significance below the level required for inclusion in the Second
EGRET Catalog (\cite{tom95}).
PKS2005$-$489 is the only XBL that has previously been observed with the 
CANGAROO 3.8m telescope albeit at a slightly higher gamma-ray energy
threshold than the observations presented here. A combined dataset containing
41 hours of on-source observations taken in 1993 and 1994 provided
an upper limit to gamma-ray emission above 2TeV of 
$1.1 \times 10^{-12} {\rm cm}^{-2}{\rm s}^{-1}$ (\cite{rob98} 1998).
In 1997 we observed PKS2005$-$489 with the 
CANGAROO telescope from the $27^{th}$
of August until the $9^{th}$ of September. The analyzable
data consist of 8 nights of observations, containing a total of $\sim 15$
hours of off-source and $\sim 17$ hours of on-source data. An analysis
of these data shows no evidence for detectable gamma-ray emission above 1.5TeV,
either in the total dataset or for any of the individual nights. The
significance of the on-source gamma-ray selected excess is $+0.01 \sigma$
with a $2 \sigma$ flux upper limit above 1.5 TeV of $7.0 \times 10^{-12}$
photons cm$^{-2}$ s$^{-1}$. 

The ASM RXTE data for PKS2005$-$489 show an average X-ray count rate
of $0.8 \pm 0.1$ for the period of CANGAROO observations. The predicted
TeV emission from this rate is $3.3 \times 10^{-12}$ photons cm$^{-2}$s$^{-1}$.
The average ASM count rate from JD2450612 to JD2450863 was $0.36 \pm 0.14$
 and the maximum count rate for any 13 day period was $0.917 \pm 0.12$.

We have made no optical observations of PKS2005$-$489.

\begin{figure}
 \picplace{70mm}
 \includegraphics{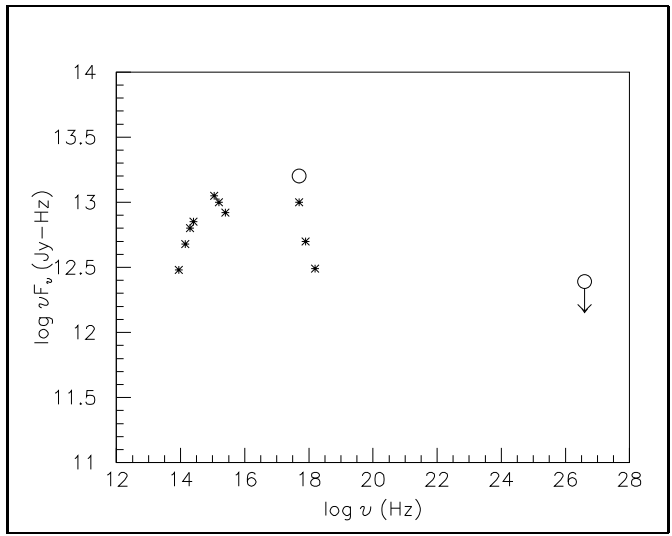}
   \caption{SED for PKS2005$-$489. The open circles show our 
measurement and the contemporaneous X-ray flux, the stars are historical (non-contemporaneous)
measurements adapted from \cite{gho95}.}
 \label{pks2005}
\end{figure}

\section{PKS2155$-$304}
PKS2155$-$304 (z=0.116) is one of the brightest BL~Lacs at X-ray energies
(\cite{lam96}), and has shown large and rapid variations in intensity
at all wavelengths (\cite{ede95} 1995). PKS2155$-$304 
was in the CGRO EGRET field a number of
times during observation phases I and II, but yielded only low
significance excesses, below the nominal $4\sigma$ EGRET source detection
limit. PKS2155$-$304 was detected, however, during a two week observation
made in November 1994. This detection shows a hard photon spectral index
of $\sim$ 1.7 at EGRET energies. As in the case of Mkn~421, the 
gamma-ray $\nu$F$_{\nu}$ SED seems to peak at energies above the EGRET energy
range (\cite{ves95} 1995).

We have made two distinct sets of CANGAROO observations 
in 1997. The first, taken
from the $27^{th}$ of September until the $8^{th}$ of October, contains
7 nights of observations with $\sim 14$ hours of off-source and 
$\sim 13$ hours of on-source data.
The second set of observations were taken to (partially) coincide with a
PKS2155$-$304 multi-wavelength campaign undertaken between the $11^{th}$
and $25^{th}$ of November 1997. A quick look analysis of EGRET data taken
in the first three and a half days of this campaign showed a detection
at a significance of $3.9 \sigma$ indicating a high source state at
GeV energies.
Unfortunately, due to the phase of the moon and poor weather, we
could not start observations until the 24th of November. We continued
observations until the $1^{st}$ of December, collecting roughly 5 hours
of both on and off-source data over 5 nights. Due to the lateness in the
year of these observations we could only observe PKS2155$-$304 for
about 1 hour each night, and at elevations between $55^{\circ}$ and
$45^{\circ}$. We estimate that the average energy threshold
of the 3.8m telescope for these observations was $\sim 2.5$ TeV.
No evidence for gamma-ray emission has been seen in 1997 observations
of PKS2155$-$304. The
significances are $+0.8 \sigma$ and $-0.2 \sigma$ for the low and high zenith
angle observations respectively, with corresponding flux limits of \\

{\noindent}F($>1.5{\rm TeV})<9.5 \times 10^{-12}$ photons ${\rm cm}^{-2}{\rm s}^{-1}$\\  
F($>2.5{\rm TeV})<6.2 \times 10^{-12}$ photons ${\rm cm}^{-2}{\rm s}^{-1}$ \\
  
The ASM count rates corresponding to the TeV observations at small and
large zenith were $0.38 \pm 0.07$ and $0.8 \pm 0.3$ respectively. The
average ASM count rate for the PKS2155$-$304 dataset is $0.38 \pm 0.02$ 
and the
peak X-ray count rate for any 12 day period was $0.9 \pm 0.1$.
Based on the X-ray flux we predict TeV gamma-ray fluxes of \\

{\noindent}F($>1.5{\rm TeV}) \sim 0.85 \times 10^{-12}$ photons ${\rm cm}^{-2}{\rm s}^{-1}$\\  
F($>2.5{\rm TeV}) \sim 0.6 \times 10^{-12}$ photons ${\rm cm}^{-2}{\rm s}^{-1}$ \\

We have contemporaneous optical observations for four nights from the
28$^{th}$ of September to the 2$^{nd}$ of October.
A clear change in the optical flux can be seen over this period, from
$36.7 \pm 0.3$ mJy to $33.4 \pm 0.4$ mJy.
The range of R band fluxes observed is comparable with other
measurements of PKS2155$-$304 (see e.g. \cite{cou95} and references therein).

\begin{figure}
 \picplace{70mm}
 \includegraphics{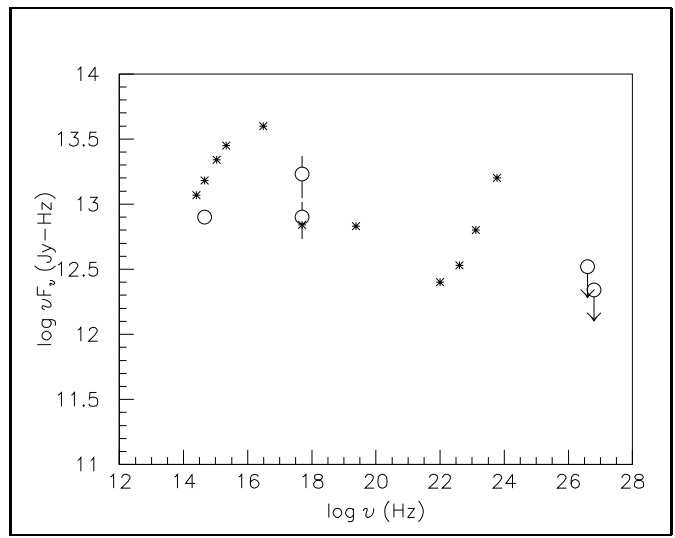}
   \caption{SED for PKS2155$-$304. Our observations and the contemporaneous
RXTE measurement are indicated by
open circles, while the non-contemporaneous historical measurements are
 indicated
by stars. The historical IR to X-ray data are adapted from \cite{ede95} (1995)
, the
CGRO OSSE data point from \cite{mcn95} and the CGRO EGRET data points
from \cite{ves95} (1995).}
 \label{pks2155}
\end{figure}

\section{Discussion}
Current knowledge about the exact location and production mechanism for
gamma-rays of energy above 300GeV in XBLs is still quite poor. 
Interpretation of observational results is also hindered
by the lack of a large catalogue of contemporaneous wideband
SED measurements for these sources. Ground based gamma-ray 
observations of XBLs have
shown that all three of the closest XBLs are observable TeV sources, 
while none of
the more distant XBLs have so far been detectable
(see e.g. \cite{cat97a,pet97}). While the statistics are
clearly very limited and biased by selection effects, it is 
suggestive that XBLs as a class do
emit detectable levels of TeV photons (at least during active phases), 
but the more distant sources are
not detectable due to photon attenuation in the IR background. 
The flux upper limits in this paper are below or roughly equal to the
predictions based on contemporaneous X-ray fluxes. 
The predictions are based on a simple model which uses the X-ray
flux to determine the expected level of TeV emission, assuming that the SED
(and hence the TeV/X-ray ratio) is the same between the known
TeV sources Mkn~421, Mkn~501 and the sources reported here.

Of particular importance for TeV gamma-ray production are 
the maximum electron energy in the jet and
the bulk Lorentz factor of the jet, which limit the maximum energy
of gamma-rays that can be produced. For PKS2155$-$304, for example, a recent
very long exposure observation by $BeppoSAX$ has been used to provide an
upper limit to the maximum energy of the electrons in the jet
($\gamma _{o} \leq 1.6 \times 10^{5}$) which rules out TeV gamma-rays
unless there is a very high degree of beaming in the source
(\cite{gio98}). It is not clear, however, how much the maximum electron 
energy increases during flaring in XBLs. The $BeppoSAX$ observation of
PKS2155$-$304, and other X-ray observations of XBLs (see e.g. \cite{gio90})
indicate spectral changes during flaring (hardening of the spectrum with
increased flux). Additionally, it has been suggested that the strong TeV 
flares seen in Mkn~421 are due to brief increases in the maximum electron
energies in the jet (\cite{mac95}).

Of the observations presented here the most promising candidate for TeV
detection is PKS2005$-$489. The RXTE results
show that we have observed this source during a relatively active phase, and
previous multi-wavelength campaigns show a SED between radio
and X-ray that is similar to Mkn~421 (\cite{sam95}). 

Optical and X-ray flux measurements indicate that PKS0548$-$322 
was at a fairly typical level of activity during the period of CANGAROO
observations. The X-ray spectrum of PKS0548$-$322 is one of the hardest of
all XBLs, although it shows considerable variation which does not
appear to be strongly correlated with the X-ray flux 
(\cite{tas95,kub98}). The lack
of correlated change in the radio spectral index indicates that the X-ray
emission is most likely synchrotron emission, with an average synchrotron
turnover frequency that is just below X-ray wavelengths and comparable to
or higher than that seen in Mkn~421 and Mkn~501. While the position of the
synchrotron turnover alone might not be a reliable indicator of TeV emission,
it is one of the defining differences between XBLs, which are known to
emit TeV gamma-rays, and RBLs, which are not. The upper limit to TeV
emission presented here does not constrain our X-ray--based prediction, but
further observations of this source by the current CANGAROO telescope, 
particularly during periods of X-ray activity, would be easily justified.
  
\section{Conclusion}
Analysis of CANGAROO observations of the three nearest southern 
hemisphere X-ray selected BL~Lacs
shows no evidence for gamma-ray emission above an energy of 1.5\,TeV. The 
$2\sigma$ upper limits to steady emission are
$4.3 \times 10^{-12} {\rm cm}^{-2}{\rm s}^{-1}$ (PKS0548$-$322),
$7.0 \times 10^{-12} {\rm cm}^{-2}{\rm s}^{-1}$ (PKS2005$-$489), and
$9.4 \times 10^{-12} {\rm cm}^{-2}{\rm s}^{-1}$ (PKS2155$-$304).
Further observations of PKS2155$-$304, taken at larger zenith angles, give
a $2\sigma$ upper limit of $6.2 \times 10^{-12} {\rm cm}^{-2}{\rm s}^{-1}$
above 2.5\,TeV. Contemporaneous X-ray and optical observations 
(for PKS0548$-$322 and PKS2155$-$304) have been used to monitor the
activity state of these sources. A simple empirical model, based on the 
measured X-ray flux, has been used to predict the level of TeV 
emission from each source. The predicted TeV fluxes are roughly equal to
or below the upper limits derived from our observations.   

\begin{acknowledgements}
This work is supported by a Grant-in-Aid in Scientific Research from the
Japan Ministry of Education, Science, Sports and Culture, and also by the 
Australian Research Council and the International Science and 
Technology Program. MDR acknowledges the receipt of a JSPS fellowship from
the Japan Ministry of Education, Science, Sport and Culture. 
The authors would like to thank Ian Glass of
the SAAO for providing the IR photometry.
PM acknowledges
the Australian Research Council for providing funding for the optical 
observatory at Woomera. 
This work has made use of the NASA-IPAC Extragalactic Data Base (NED).
\end{acknowledgements}

\end{document}